\documentclass[pra,twocolumn,amsmath,amssymb,superscriptaddress]{revtex4-1} %
\usepackage{epsfig}
\usepackage{subfigure}
\usepackage{graphicx}
\usepackage{dcolumn}
\usepackage{stmaryrd}
\usepackage{mathrsfs}
\usepackage{pifont}
\usepackage{amsthm}
\usepackage{amssymb}
\usepackage{amsbsy}
\usepackage{amsmath}
\usepackage{bm}
\usepackage{latexsym}
\usepackage{color}
\usepackage[usenames,dvipsnames]{xcolor}
\usepackage[colorlinks=true,citecolor=Cerulean,linkcolor=RubineRed,urlcolor=Cerulean]{hyperref}

\newcommand{\beq}{\begin{equation}}
\newcommand{\eeq}{\end{equation}}
\newcommand{\beqa}{\begin{eqnarray}}
\newcommand{\eeqa}{\end{eqnarray}}
\newcommand{\lam}{\lambda}
\newcommand{\De}{\Delta}

\newcommand{\da}{\dagger}
\newcommand{\non}{\nonumber}

\newcommand{\la}{\langle}
\newcommand{\ra}{\rangle}

\newcommand{\ga}{\gamma}
\newcommand{\Ga}{\Gamma}

\newcommand{\pa}{\partial}
\newcommand{\si}{\sigma}

\newcommand{\al}{\alpha}
\newcommand{\para}{\|}

\def\pra#1{{ Phys.\ Rev. A\/} {\bf#1}}

\def\prl#1{{ Phys.\ Rev.\ Lett.} {\bf#1}}

\def\pla#1{{ Phys.\ Lett. A\/} {\bf#1}}
\def\rmp#1{{ Rev. \ Mod. \ Phys.} {\bf#1}}
\def\nat#1{{ Nature} {\bf#1}}

\begin{document}

\title{Fundamental Speed Limits to the Generation of Quantumness}

\author{Jun Jing}
\affiliation{Institute of Atomic and Molecular Physics \\and Jilin Provincial Key Laboratory of Applied Atomic and Molecular Spectroscopy, Jilin University, Changchun 130012, Jilin, China}
\affiliation{Department of Theoretical Physics and History of Science, The University of the Basque Country (EHU/UPV), PO Box 644, 48080 Bilbao, Spain}

\author{Lian-Ao Wu}
\affiliation{Department of Theoretical Physics and History of Science, The University of the Basque Country (EHU/UPV), PO Box 644, 48080 Bilbao, Spain}
\affiliation{Ikerbasque, Basque Foundation for Science, 48011 Bilbao, Spain}

\author{Adolfo del Campo}
\affiliation{Department of Physics, University of Massachusetts, Boston, MA 02125, USA}

\date{\today}

\begin{abstract}
Quantum physics dictates fundamental speed limits during time evolution. We present a quantum speed limit governing the generation of nonclassicality and the mutual incompatibility of two states connected by time evolution. This result is used to characterize the timescale required to generate a given amount of quantumness under an arbitrary physical process. The bound is shown to be tight under pure dephasing dynamics. More generally, our analysis reveals the dependence on the initial state and non-Markovian effects.
\end{abstract}

\date{\today}

\maketitle

\section{Introduction}

Quantum speed limits (QSLs) provide a lower bound to the rate at which a physical system can evolve. Due to their fundamental nature, QSLs have found applications in a wide variety of fields including quantum information processing~\cite{Lloyd00}, quantum metrology~\cite{rafal,GLM11}, quantum simulation~\cite{DiCandia15}, quantum thermodynamics~\cite{delCampo14}, quantum critical dynamics~\cite{Rezakhani10,DRZ12}, quantum control~\cite{DR08,Caneva09,Hegerfeldt13,SS15} and other quantum technologies.

The first rigorous QSL was derived as a time-energy uncertainty relation providing a lower bound to the required passage time $\tau$ for a system to evolve from an initial state $|\psi_0\ra$ to a final state $|\psi_\tau\ra=U(\tau,0)|\psi_0\ra$, where $U(\tau,0)$ is the time-evolution operator associated with the driving Hamiltonian $H$. It was shown that $\tau\geq\arccos(|\la\psi_0|\psi_\tau\ra|)/\De E$, where $\De E$ is the energy dispersion of the initial state~\cite{MT45,Fleming73,Bhattacharyya83,Vaidman92,Uhlmann92,Pfeifer93}. The modern formulation of QSL for unitary processes takes into account an alternative expression as an upper bound for the speed of evolution, the mean energy of the system, that can replace the role of energy dispersion $\De E$~\cite{ML98,Lloyd00,LT09}. A geometric interpretation provides an intuitive understanding of the QSL bound as a brachistochrone where the geodesic set by the Fubini-Study metric in (projective) Hilbert space is travelled at the maximum speed of evolution achievable under a given Hamiltonian dynamics~\cite{AA90,Brody03,Russell14,Russell15}. Time-optimal evolutions are often explored in the context of quantum control theory, where the existence of a QSL has been shown to limit the performance of algorithms aimed at identifying optimal driving protocols~\cite{Caneva09,Hegerfeldt13}. More recently, QSLs have been extended to open quantum dynamics where the system of interest is embedded in an environment~\cite{QSLopen1,QSLopen2,QSLopen3,QSLopen4,QSLopen5}. The evolution needs not be restricted to a master equation and can be alternatively described by general quantum channels~\cite{QSLopen1,QSLopen2}. These new QSLs to non-unitary evolution have been formulated in terms of a variety of norms of the generator of the dynamics. Similar bounds can be expected to apply to classical processes as well~\cite{Flynn14}.

While for certain applications it might suffice to characterize QSL exclusively through the properties of the generator of the dynamics~\cite{rafal,Uzdin14}, a reference to the initial and time-evolving states generally becomes unavoidable. This is particularly the case for externally driven systems or open quantum systems exhibiting non-Markovian effects resulting from the finite-memory of the environment. We further notice that when the dynamics of a system is registered by monitoring a given observable, the standard QSL governing the fidelity decay can become too conservative, and even fail to capture the right scaling of the time scale of interest with the system parameters. A prominent example is provided by thermalization, where the identification of the relevant time scale remains an open problem~\cite{Eisert15}.

Identifying the minimal passage time for arbitrary physical processes is as well crucial to understanding the quantum-to-classical transition~\cite{Zurek}. This transition is of particular relevance in composite quantum systems exhibiting non-classical correlations, with applications to a variety of fields~\cite{Modi12}. To characterize the crossover between the quantum and classical worlds of a single physical system, one can define the notion of quantumness on the non-commutativity of the algebra of observables~\cite{quantumness1} in a way that it is experimentally measurable~\cite{quantumness2}. In this framework a system is found to be classical if all its accessible states commute with each other.

In this work, we exploit the definition of quantumness involving the non-commutativity of the initial and final states of the system of interest. We derive lower bound for the timescale required to generate a given amount of quantumness, that quantifies the degree to which the time-evolving state is mutually incompatible with the initial state under arbitrary dynamics. The new bound allows one to classify different dynamics according to their power to generate nonclassicality and is shown to be saturated under pure dephasing dynamics, whether it is induced by a Markovian or a non-Markovian environment.

\section{Quantum speed limit to the dynamics of quantumness}

The nonclassicality of quantum systems can be conveniently quantified using the Hilbert-Schmidt norm of the commutator of two states, which is proposed to witness the ``state incompatibility" between any two admissible states $\rho_a$ and $\rho_b$~\cite{quantumness1,quantumness2}. The ``quantumness'' is then defined as
\beqa
\label{Qab}
Q(\rho_a, \rho_b)&\equiv&2\para[\rho_a, \rho_b]\para^2\non\\
&=&-4{\rm Tr}\left[(\rho_a\rho_b)^2-\rho_a^2\rho_b^2\right],
\eeqa
where the pre-factor is required for normalization and $\para A\para^2\equiv {\rm Tr}(A^\da A)$ is the Hilbert-Schmidt norm of $A$. As a quantumness witness,
\beqa
0\leq Q(\rho_a, \rho_b)\leq 1
\eeqa
and $Q(\rho_a, \rho_b)=0$ iff $[\rho_a, \rho_b]=0$~\cite{quantumness1,quantumness2}. Choosing $\rho_a=\rho_0$ and $\rho_b=\rho_t$, $Q(\rho_0, \rho_t)$ allows one to quantify the capacity of an arbitrary physical process to generate or sustain quantumness in case of $[\rho_0, \rho_t]\neq0$. Clearly, if $\rho_0$ is a diagonal density matrix in a given basis and time evolution just alters the weight distribution without generating coherences, the quantumness between the initial and time-evolving states $Q(\rho_0, \rho_t)$ remains zero. Consequently we can generally expect a QSL different in nature from those previously derived for the fidelity decay, which would remain valid as weaker lower bounds. The connection between different QSLs will be made explicit below.

We consider the time-evolution of the initial density matrix to be described by a master equation of the form
\beqa
\dot{\rho}_t=\mathcal{L}\rho_t,
\eeqa
where $\mathcal{L}$ is the Louville super-operator. The rate at which quantumness can vary is then exactly given by
\beqa
\dot{Q}(\rho_0, \rho_t)=-4{\rm Tr}([\rho_0, \rho_t][\rho_0, \mathcal{L}\rho_t]).
\eeqa
As an example, $\mathcal{L}\rho_t=-i[H, \rho_t]/\hbar$ for unitary dynamics, i.e., in a closed system. Using the Cauchy-Schwarz inequality, i.e., $|{\rm Tr}(A^\dag B)|\leq \|A\| \|B\|$ and by virtue of $\sqrt{Q}=\sqrt{2}\para[\rho_0, \rho_t]\para$, it follows from the definition of quantumness in Eq.~(\ref{Qab}) that
 \beqa
 |\dot{Q}(\rho_0, \rho_t)|\leq2\sqrt{2Q}\para[\rho_0, \mathcal{L}\rho_t]\para.
 \eeqa
To derive a quantum speed limit we integrate from $t=0$ to $t=\tau$. Note that $Q(\rho_0, \rho_0)=0$, and $\int_0^{\tau}dt|\dot{Q}|/\sqrt{Q}\geq\left|\int_0^Q dQ'/\sqrt{Q'}\right|=2\sqrt{Q(\rho_0, \rho_\tau)}$. As an upshot, the time in which quantumness can emerge is lower-bounded by
\beqa \label{QSL1}
\tau\geq\tau_{Q}\equiv\sqrt{\frac{Q(\rho_0, \rho_\tau)}{2}}\frac{1}{\overline{\para[\rho_0, \mathcal{L}\rho_t]\para}},
\eeqa
where the time-average is denoted by $\bar{X}=\tau^{-1}\int_0^\tau Xdt$. We note that even if $\mathcal{L}$ is explicitly time-independent, i.e., the parameters in the equation of motion are constants, then $\overline{\para[\rho_0, \mathcal{L}\rho_t]\para}$ can {\it not} be reduced to $\para[\rho_0, \mathcal{L}\rho_\tau]\para$ since $\rho_t$ is a function of time.

It is worth pointing out that Eq.~(\ref{QSL1}) suggests
\beqa
\overline{\para[\rho_0, \mathcal{L}\rho_t]\para}
\eeqa
as an upper bound for the speed of evolution of quantumness. Clearly, this quantity can be further upper bounded using the triangular and Cauchy-Schwarz inequalities by $2\overline{\para\mathcal{L}\rho_t\rho_0\para}$. The resulting bound closely resembles the QSL derived by studying the reduced dynamics of an open quantum system in terms of the fidelity decay~\cite{QSLopen2,QSLopen3,QSLopen4}. We note however that this bound is more conservative than that given by $\tau_{Q}$ in Eq.~(\ref{QSL1}). Weaker bounds could be derived as well exploiting the fact that $\overline{\para\mathcal{L}\rho_t\rho_0\para}\leq
\overline{\para\mathcal{L}\rho_t\para\para\rho_0\para}
\leq\overline{\para\mathcal{L}\rho_t\para}$, or conversely $\overline{\para\mathcal{L}\rho_t\rho_0\para}\leq
\overline{\para\mathcal{L}^{\dag}\rho_0\para\para\rho_t\para}\leq
\overline{\para\mathcal{L}^{\dag}\rho_0\para}$ using the adjoint of the generator $\mathcal{L}^{\dag}$~\cite{QSLopen2}. We shall not pursue this goal here.

\section{The lower bound for quantum dynamics}

The lower bound obtained in Eq.~(\ref{QSL1}) constitutes the main result of this work. In what follows, this bound is analyzed in a series of relevant scenarios, that will be used to identify the salient physical principles governing the generation of quantumness. After a discussion of its dynamics in isolated systems we consider a system embedded in an environment, exhibiting possible non-Markovian effects, and discuss the limits of pure dephasing and dissipative processes.

\subsection{Unitary quantum dynamics}

Consider a general two-parameter unitary transformation for a two-level system (setting $\hbar\equiv1$ from now on)
\beqa
U=\cos\theta+i\sin\theta(\si_x\cos\al+\si_y\sin\al),
\eeqa
where $\theta$ and $\al$ are arbitrary real functions of time and $\si$ is the Pauli operator~\cite{Jing}. When the system is prepared in an initially pure states $\rho_0=|0\ra\la0|$, it evolves into $\rho_t=|\psi_t\ra\la\psi_t|$ with $|\psi_t\ra=\sin\theta|1\ra-ie^{i\alpha}\cos\theta|0\ra$. In this case, the system Hamiltonian is found to be
 \beqa
 H&=&i\dot{U}U^\da,\\
 &=&\si_x(-\dot{\theta}\cos\alpha+\dot{\alpha}\sin\theta\cos\theta\sin\alpha)
\non\\
& & -\si_y(\dot{\theta}\sin\alpha+\dot{\alpha}\sin\theta\cos\theta\cos\alpha)+\si_z\dot{\alpha}\sin^2\theta. \non
\eeqa
For the creation of quantumness, we require $[\rho_0, \rho_t]\neq0$, i.e., $\sin2\theta\neq0$. It follows from Eq.~(\ref{QSL1}), that
\beqa
\tau\geq\tau_{Q}=\frac{|\sin2\theta|}{\sqrt{2}\overline{\sqrt{X}}},
\eeqa
where
\beqa\non 
X&=&-2\dot{\alpha}\sin^2\theta\sin4\theta(\dot{\alpha}\cos^2\alpha\sin\theta
+\dot{\theta}\sin2\alpha)\\  &+&2\dot{\theta}^2\cos^22\theta+\dot{\alpha}^2\sin^2\theta.
\eeqa
In case that both $\dot{\theta}$ and $\alpha$ are constant numbers, the dynamics is generated by $H=-\dot{\theta}(\si_x\cos\alpha+\si_y\sin\alpha)$, and the exact evolution time saturates the lower bound $\tau=\tau_Q$, for $\theta<\pi/4$. It is worth pointing out that in this regime of $\theta$, $\tau_Q$ is independent of the angle $\alpha$. Obviously, this lower bound expression is invalid when $\theta$ becomes $\pi/4$. However, it is easy to find that whenever $\dot{\alpha}$ is a constant number, then $\tau_Q(\theta=\pi/4)=\tau/|\alpha|$. Therefore, the QSL ruling the evolution of quantumness exhibits a pronounced dependence on the initial and final states.

A similar analysis can be extended to higher-dimensional systems. Consider the stimulated Raman adiabatic passage (STIRAP)~\cite{STIRAP} in a three-level atomic system, under the Hamiltonian as 
\begin{equation}
H(t)=i\left(\begin{array}{ccc}
0 & \dot{\al}\cos\theta & -\dot{\theta} \\
-\dot{\al}\cos\theta & 0 & -\dot{\al}\sin\theta \\
\dot{\theta} & \dot{\al}\sin\theta & 0
\end{array}\right).
\end{equation}
The system can be transferred from $\rho_0$ to $\rho_\tau=|\psi_\tau\ra\la\psi_\tau|$, where $|\psi_\tau\ra=-\sin\theta(\tau)|2\ra+\cos\theta(\tau)|0\ra$ without disturbing the quasistable state $|1\ra$. The QSL bound becomes tight and matches the exact time of evolution $\tau_Q=\tau$ when $\theta<\pi/4$ and $\dot{\theta}$ and $\alpha$ are time-independent.

\subsection{Nonunitary process}

In this subsection, we consider the scenario of open quantum systems. We will use the quantum-state-diffusion (QSD) equation~\cite{QSD1,QSD2} as a general framework to derive the exact master equation before discussing the relevant QSL. In doing so, we treat both Markovian and non-Markovian environments in a unified way. In particular, we consider an Ornstein-Uhlenbeck process for the environmental noise. The correlation function reads
\beqa
G(t,s)=\frac{\Ga\ga}{2}e^{-\ga|t-s|},
\eeqa
where $0<\ga<\infty$ and the lower and upper limits of $\ga$ correspond to the strongly non-Markovian and Markov environments, respectively. For a single two-level system, the system-environment Hamiltonian is
 \beqa
 H_{\rm tot}=\frac{\omega}{2}\sigma_z+\sum_{\lam}(g_{\lam}La_{\lam}^\da+h.c.)
+\sum_{\lam}\omega_{\lam}a_{\lam}^\da a_{\lam},
\eeqa 
where $L$ is the coupling operator and $a_\lam$ ($a^\da_\lam$) is the annihilation (creation) operator for the $\lam$-th environmental mode.

\begin{figure}[t]
\centering
\includegraphics[height=0.6\linewidth]{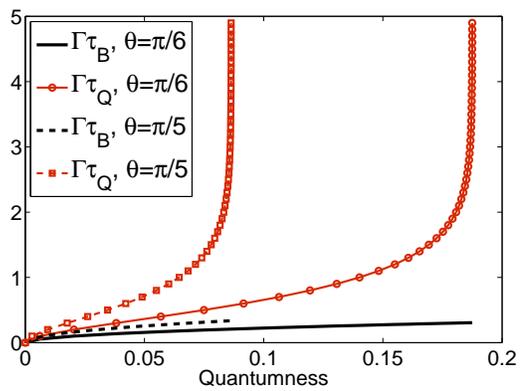}
\caption{(Color online) The quantum speed limit timescales $\tau_{Q}$ (based on quantumness) and $\tau_B$ (based on the fidelity) as a function of quantumness $Q$ in the Markovian pure-dephasing processes with different initial states: $\rho_0=|\psi_0\ra\la\psi_0|$, where $|\psi_0\ra=\cos\theta|1\ra+\sin\theta|0\ra$. Under pure dephasing the bound $\tau_{Q}$ is shown to be identical to the exact time $\tau$ in which quantumness is generated.}
\label{Dep}
\end{figure}

When $L=\si_z$, QSD equation describes pure dephasing process, and in the rotating frame with respect to the system bare Hamiltonian, the exact super-operator $\mathcal{L}$ is found to be
\beqa
\mathcal{L}\rho_t=[\bar{G}(t)+\bar{G}^*(t)](\sigma_z\rho_t\sigma_z-\rho_t),
\eeqa
where $\bar{G}(t)=\int_0^tdsG(t,s)$. If $\rho_0=|\psi_0\ra\la\psi_0|$ where $|\psi_0\ra=\cos\theta|1\ra+\sin\theta|0\ra$, then $\la0|\rho_\tau|0\ra=\la0|\rho_0|0\ra$, $\la1|\rho_\tau|1\ra=\la1|\rho_0|1\ra$, and $\la1|\rho_\tau|0\ra=\sin\theta\cos\theta e^{-\beta}$, where $\beta$ is a positive number. Here $\beta\equiv2F(\tau)=2\int_0^\tau dtf(t)$ and $f(t)\equiv\bar{G}(t)+\bar{G}^*(t)$. In the Markov limit, $f(t)\rightarrow\Ga$ and then $F(t)=\Ga t$. After substituting $\rho_0$, $\rho_\tau$ and $\mathcal{L}\rho_t$ into Eq.~(\ref{QSL1}), it is found that
\begin{eqnarray}\label{Qt}
Q(\rho_0,\rho_{\tau})&=&\frac{1}{4}\sin^24\theta(1-e^{-\beta_\tau})^2, \\ \label{QSLde}
\tau&=&\tau_{Q}=-\frac{1}{2\Ga}\ln\left(1-\frac{2\sqrt{Q}}{|\sin4\theta|}\right),
\end{eqnarray}
where $\beta_\tau\equiv2\Ga\tau$. Remarkably, the bound is tight and reachable under pure-dephasing dynamics, when $\tau=\tau_{Q}$ as shown in Fig.~\ref{Dep}. Equation~(\ref{QSLde}) also applies to the non-Markovian case as long as $\beta_\tau$ in Eq.~(\ref{Qt}) is modified into $2\int_0^\tau f(t)dt$. Qualitatively, QSL timescale depends on the choice of initial state parameter $\theta$, specifically, the initial population distribution that is determined by $\cos(2\theta)$. QSL is therefore symmetric as a function of $\theta$ with respect to $\theta=\pi/4$.

In Fig.~\ref{Dep}, for different initial states, we compare the new QSL timescale $\tau_Q$ obtained in Eq.~(\ref{QSL1}) and that $\tau_B$ based on the fidelity evolving with time (see Ref.~\cite{QSLopen2}) in presence of a Markovian dephasing environment. Specifically, it was then shown that for an initially pure state, the minimum time for the (squared) fidelity or relative purity $f(t)={\rm Tr}[\rho_0\rho_t]$ to decay to a given value $f(\tau)$ is lower bounded by $\tau_B=|1-f(\tau)|/\|\mathcal{L}(\rho_0)\|$ whenever the dynamics is governed by a master equation of the form $d_t\rho_t=\mathcal{L}(\rho_t)$. Figure~\ref{Dep} illustrates $\tau_Q$ not only provides a tighter bound than $\tau_B$ for the generation of quantumness, but also it actually captures the real evolution pattern. Furthermore, it is known that as the system progressively goes to a steady state, which depends on the initial coherence between the up and down states, the dephasing rate should be asymptotically slowed down. This pattern has not been captured by $\tau_B$. When the quantumness approaches a final value determined by $\theta$, $\tau_Q$ increases rapidly while the rate of $\tau_B$ is nearly invariant.

\begin{figure}[t]
\centering
\includegraphics[height=0.6\linewidth]{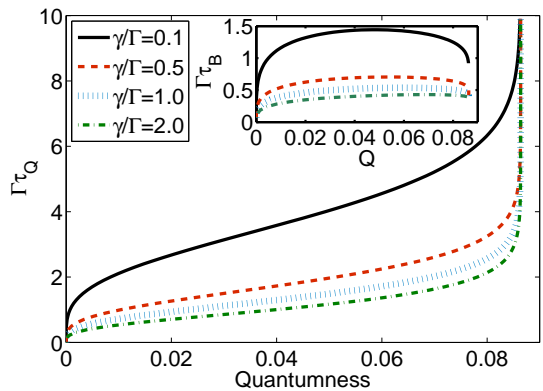}
\caption{(Color online) Dependence of the quantum speed limit timescale $\tau_Q$ on the memory parameter $\ga$ in the non-Markovian pure-dephasing dynamics as a function of quantumness $Q$ ($\theta=\pi/5$). The inset shows the bound $\tau_B$ derived from the fidelity decay, which fails to capture the correct behavior. }\label{Dep2}
\end{figure}

Next we consider the effect of the environmental memory, which is parameterized by $\ga$, on the QSL timescale. In Fig.~\ref{Dep2}, $\tau_Q$ is evaluated for a fixed initial state (with $\theta=\pi/5$) and the other parameters except $\ga$ and $Q$. The dependence of $\tau_Q$ on the quantumness $Q$ of system and environment is monotonic. The environmental memory timescale is inversely proportional to $\ga$. As an upshot, in presence of a strongly non-Markovian environment the evolution speed is greatly suppressed, resulting in larger values of $\tau_Q$. Yet it is found that at the end of the system dephasing, the quantum speed limit timescale quickly approaches the same asymptotical value. The difference between the QSL timescale of the system in the extremely non-Markovian environment [$\tau_Q(\ga/\Ga=0.1)$] and that in a nearly Markov environment [$\tau_Q(\ga/\Ga=2.0)$] is increased with increasing $Q$ before the system goes to the steady state.

For an $n$-qubit system in a common dephasing environment, we can rigorously discuss the scaling behavior of QSL for certain states. By a treatment in the Kraus representation~\cite{nqDep}, a general GHZ state $|\psi_0\ra=\cos\theta|1^{\otimes n}\ra+\sin\theta|0^{\otimes n}\ra$ evolves into $\rho_t=C(t)\circ\rho_0$. Here $\circ$ denotes the entry-wise product and effectively $C(t)$ (as well as $\rho_t$) can be expressed in a $2\times2$ matrix expanded by $|0^{\otimes n}\ra$ and $|1^{\otimes n}\ra$, where the off-diagonal terms are $r^{n^2}$ with $r=e^{-\beta}$ and the diagonal terms are unity. By Eqs.~(\ref{Qt}) and (\ref{QSLde}), we can find that when $\beta$ is sufficiently small (e.g., with a Markov environment, $\beta=2\Ga t$ is small in the short time limit), both the quantumness $Q$ and QSL time $\tau_Q$ scale with the number of qubits $n$ as $n^2$.

When $L=\si_-$, the total Hamiltonian describes a dissipation (energy relaxation) model, whose exact super-operation $\mathcal{L}$ is found to be
\beqa
\mathcal{L}\rho_t=P(t)[\sigma_-\rho_t, \sigma_+]+h.c.,
\eeqa
where $P(t)=\int_0^tdsG(t,s)p(t,s)$ and $\pa_tp(t,s)=P(t)p(t,s)$ with $p(s,s)=1$. Starting from the same pure state as above, in the dissipation model, the time-evolving density matrix satisfies $\la1|\rho_\tau|1\ra=\cos^2\theta e^{-\xi-\xi^*}$ and $\la1|\rho_\tau|0\ra=\sin\theta\cos\theta e^{-\xi}$, where $\xi=\bar{P}(\tau)\equiv\int_0^{\tau}dtP(t)$ is a complex function of time. In the Markov limit, $P(t)=\Ga/2$ and then $\xi=\Ga\tau/2$. In the non-Markovian situation, $P(t)$ satisfies $\pa_tP(t)=\Ga\ga/2-\ga P(t)+P^2(t)$ with $P(0)=0$. Consequently, according to Eq.~(\ref{QSL1}), it is found that
\begin{eqnarray}
Q(\rho_0,\rho_{\tau})&=&\sin^22\theta|1-2e^{-2b}\cos^2\theta+e^{-b-ic}\cos2\theta|^2 \non\\ \label{Qt2} &+&\sin^42\theta\sin^2ce^{-2b},
\\
 \para[\rho_0, \mathcal{L}\rho_t]\para^2&=&\frac{\sin^22\theta}{2}[|d\sin2\theta|^2+|P(t)e^{-b(t)-ic(t)}\cos2\theta \non\\ \label{para2} &-&2e^{-2b(t)}[P(t)+P^*(t)]\cos^2\theta|^2],
\end{eqnarray}
where $b\equiv b(\tau)={\rm Re}[\bar{P}(\tau)]$, $c\equiv c(\tau)={\rm Im}[\bar{P}(\tau)]$, and $d\equiv d(t)={\rm Im}[P(t)e^{-\xi(t)}]$. Note here $\sin2\theta$ is not allowed to be zero, otherwise, $Q(\rho_0, \rho_\tau)$ will vanish according to its definition in Eq.~(\ref{Qab}). Equations~(\ref{Qt2}) and (\ref{para2}) indicate that in the dissipation model, it is hardly to find a closed analytical expression for $\tau_Q$, and one has to resort to the numerical evaluation.

\begin{figure}[t]
\centering
\includegraphics[width=0.8\linewidth]{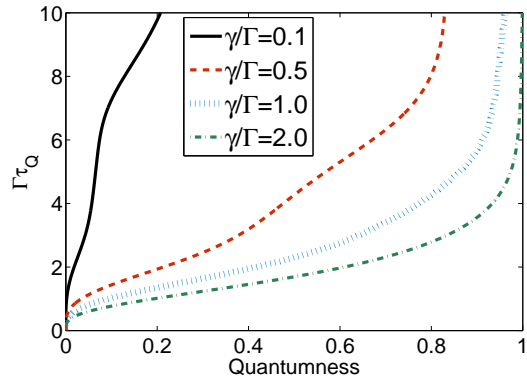}
\caption{(Color online) Dependence of the quantum speed limit timescale on the memory parameter $\ga$ of the non-Markovian dissipative process as a function of the quantumness $Q$, for $\theta=\pi/4$.}\label{Diss}
\end{figure}

In Fig.~\ref{Diss}, we demonstrate the dependence of the QSL timescale on the environmental memory parameter $\ga$, measured in units of  the system-environment coupling strength $\Ga$,  for a fixed initial state. From the numerically exact dynamics, we find that  $\tau_Q$ monotonically decreases with increasing $\ga$ and that the case with a nearly Markovian environment (see e.g., the dot-dashed line for $\ga/\Ga=2.0$), $\tau_Q$ approaches a steady value. As expected in an environment with short memory time, the energy dissipated into the environment from the system has nearly no chance to come back to the system. The dissipation process becomes therefore irreversible. This favors the evolution of the system towards a final incompatible state. As a result, two different regimes are observed. For nearly memoryless dynamics, $\ga/\Ga\geq1$, the QSL timescale is found to rapidly increase as the system approaches the steady state through an approximate exponential decay. Regarding the spectral function $G(t,s)$, a smaller $\ga$ then yields a lesser damping rate of the system. In the strong non-Markovian regime $0.1\leq\ga/\Ga<1$, the pattern becomes complex and the QSL timescale appears to be greatly enhanced by decreasing $\ga$. In this regime, it is difficult for the time-evolving state to become classically incompatible with the initial state. 

\section{Conclusion}

We have studied the generation of nonclassicality via the quantumness witness defined as the Hilbert-Schmidt norm of the commutator of the initial and the final quantum states, resulting from time evolution. For arbitrary physical processes we have derived a quantum speed limit that sets the minimum timescale $\tau_Q$ for the generation of a given amount of  quantumness. This novel QSL has been computed and analyzed in a variety of relevant scenarios including unitary evolution, pure dephasing, and energy dissipation. In addition, we have discussed the generation of quantumness in  non-unitary evolutions, by employing exact quantum-state-diffusion equations.

While standard quantum speed limits characterizing the fidelity decay become too conservative and even fail to capture the correct dependence of this timescale on the parameters of the system, the new bound is tight and can be saturated under pure dephasing dynamics, whether induced by a Markovian or non-Markovian environment.

\section*{Acknowledgments}

It is a pleasure to thank M. Beau and I. L. Egusquiza for discussions and a careful reading of the manuscript. We acknowledge grant support from the Basque Government (grant IT472-10), the Spanish MICINN (No. FIS2012-36673-C03-03), the National Science Foundation of China Nos. 11575071, and Science and Technology Development Program of Jilin Province of China (20150519021JH). Funding support from UMass Boston (project P20150000029279) and the John Templeton Foundation is further acknowledged.

\end{document}